\documentclass[aps,prl,reprint]{revtex4-2}

\usepackage{bm, amsmath, amsfonts, mathtools, physics}
\usepackage[dvipdfmx]{graphicx}
\usepackage{ulem}
\usepackage{here}

\usepackage{color}

\begin{document}

   \title{Attraction-Induced Cluster Fragmentation and Local Alignment \\ in Active Particle Systems}
  
  \author{Sota Shimamura${}^{1}$}
    \email{shimamura-sota593@g.ecc.u-tokyo.ac.jp}

  \author{Nen Saito${}^{2,3}$}
  
  \author{Shuji Ishihara${}^{1,4}$}

  \affiliation{
  ${}^1$~Graduate School of Arts and Sciences, The University of Tokyo, Komaba 3-8-1, Meguro-ku, Tokyo, 153-8902, Japan \\
  ${}^2$~Graduate School of Integrated Science for Life, Hiroshima University, 1-3-1 Kagamiyama, Higashi-Hiroshima City, Hiroshima, 739-8526, Japan \\
  ${}^3$~Exploratory Research Center on Life and Living Systems (ExCELLS), National Institutes of Natural Sciences, Aichi, Japan \\
  ${}^4$~Universal Biology Institute, The University of Tokyo, Tokyo, Japan
  }

\begin{abstract}
We numerically studied active Brownian particles with attractive interactions.
Contrary to our intuition, the attractive force between particles disrupts the formation of a single cluster observed in motility-induced phase separation, giving rise to a multi-cluster state characterized by a power-law distribution of cluster sizes. 
Remarkably, the self-propulsion directions spontaneously align within each cluster, resulting in enhanced cluster motility despite the absence of alignment interactions.
This study revealed the intricate role of attractive interactions in the aggregation of motile systems.
\end{abstract}

\maketitle
Attractive interactions play a fundamental role in organizing the collective behaviors of motile systems. 
For example, epithelial cells can detach from the parent tissue and migrate as cohesive groups~\cite{Friedl2009, Vedula2013, Lu2021}.
Similarly, bacterial swarms can fragment into transient clusters that form and scatter dynamically~\cite{Copeland2009,Zhang2010,Peruani2012}.
These dynamics are governed by a delicate balance between the internal impetus of individuals to explore and the adhesive force that binds them.
The theoretical framework of active matter systems, which describes how local interactions among motile particles can produce large-scale structures, offers a unified perspective for understanding such collective behavior~\cite{Fodor2018,Shaebani2020}.
Recent studies have shown that attraction is not the only mechanism leading to cluster formation.
Active Brownian particles (ABPs), which propel themselves in random directions with a finite persistence time, exhibit spontaneous aggregation through excluded-volume interactions alone, eventually forming a single large cluster. This phenomenon is known as motility-induced phase separation (MIPS)~\cite{Fily2012, Cates2015}.  
Given that cluster formation can occur even in the absence of attractive interactions, the role of attraction in active particles remains unclear.
Do attractive interactions simply enhance cluster formation in MIPS?
How do the dynamic processes of cluster formation and fragmentation emerge from the interplay between attractive forces and self-propulsion?

An effective approach to answering these questions is to study ABPs with attractive interactions and analyze how both attraction and motility influence the system's behavior.
However, most previous studies have focused on ABPs with only repulsive interactions, and only a few have addressed the effects of attractive interactions.
For instance, Sarker et al. introduced an ABP system with an attractive interaction potential featuring a broad flat minimum and demonstrated that a fluid-like cluster reminiscent of cell colonies can emerge~\cite{Sarkar2021}. 
Caprini and L\"owen reported that attractive ABPs exhibit a global velocity alignment within an attraction-mediated cluster~\cite{Caprini2023}.
These two studies show that ABPs with attractive forces can incompletely aggregate into multiple clusters, as also observed in both experimental and theoretical studies~\cite{Theurkauff2012, SchwarzLinek2012, Mognetti2013}. However, the nature of the multi-cluster state remains largely unexplored, and it is still unclear how MIPS, attraction-driven aggregation, and the multi-cluster states evolve and transition as the attraction strength is varied.

In this study, we investigated an ABP model with attractive forces, focusing on the multi-cluster state.
This state is particularly interesting, not only because it emerges from the competition between motility and attractive interactions, but also because it closely resembles the cellular cluster formation observed in various biological experiments~\cite{Mendes2001, Zhang2010, Peruani2012, Chen2012}.
We identified several intriguing properties of this state, including a power-law distribution of cluster sizes and an enhanced cluster velocity driven by the cooperative alignment of particle self-propulsion directions within each cluster despite the absence of explicit alignment interactions.

\textit{Model:} We consider a two dimensional system consisting of $N$ self-propelled particles that interact with each other through attractive potential. 
The position of the $i$-th particle $\bm{r}_i$ and its direction $\theta_i$ obey the following Langevin equations,
\begin{align}
    \label{eq:EqR}
    &m\ddot{\bm r_i} = -\gamma \dot{\bm r_i} +f\bm n_i + \bm{F}_i  + \sqrt{2\gamma k_{\mathrm B} T} \bm \xi_i~,\\
    \label{eq:EqTheta}
    &~~~~~~~~~~~~~~~~\dot {\theta_i}  =  \sqrt{2/\tau_{\theta}} \eta_i(t)~.
\end{align}
Here, $-\gamma \dot{\bm r_i}$ denotes the drag force. 
We consider a situation in which drag dominates over inertia, and the system is nearly in an overdamped regime. 
$f$ is the strength of the self-propelled force and $\bm n_i = (\cos\theta_i,\sin\theta_i)$ is a unit vector representing the direction of the self-propulsion force. 
The direction $\theta_i$ varies over time and is driven only by noise, as indicated by Eq.~\eqref{eq:EqTheta}, where $\eta_i(t)$ represents the normalized Gaussian white noise satisfying $\langle \eta_i(t) \rangle=0$ and $\langle \eta_i(t)\eta_j(t') \rangle = \delta_{ij} \delta(t-t')$. 
Note that the directions $\theta_i$ are independent and not correlated with each other in the ABP model.
The P\'eclet number, a dimensionless parameter to be controlled below, is defined by $\mathrm{Pe}=\tau_\theta/\tau_R$, where $\tau_R=\gamma\sigma/f$ represents the time required for a single particle to travel a unit distance $\sigma$ with its self-propelled force. 
$\mathrm{Pe}$ indicates the relative strength of self-propulsion against the particle diffusivity determined by the direction change~\cite{Cates2015}.
The last term in Eq.~\eqref{eq:EqR} represents noise in the positional dynamics, which is characterized by the effective temperature $T$ and normalized Gaussian white noise $\bm{\xi}_i(t)=(\xi^x_i,\xi^y_i)$, whose components satisfy $\langle \xi_i(t) \rangle=0$ and $\langle \xi_i(t)\xi_j(t') \rangle = \delta_{ij} \delta(t-t')$.
We assume that this noise term is sufficiently small.

\begin{figure}[tbhp]
    \includegraphics[width=\columnwidth,bb=0 0 850 610]{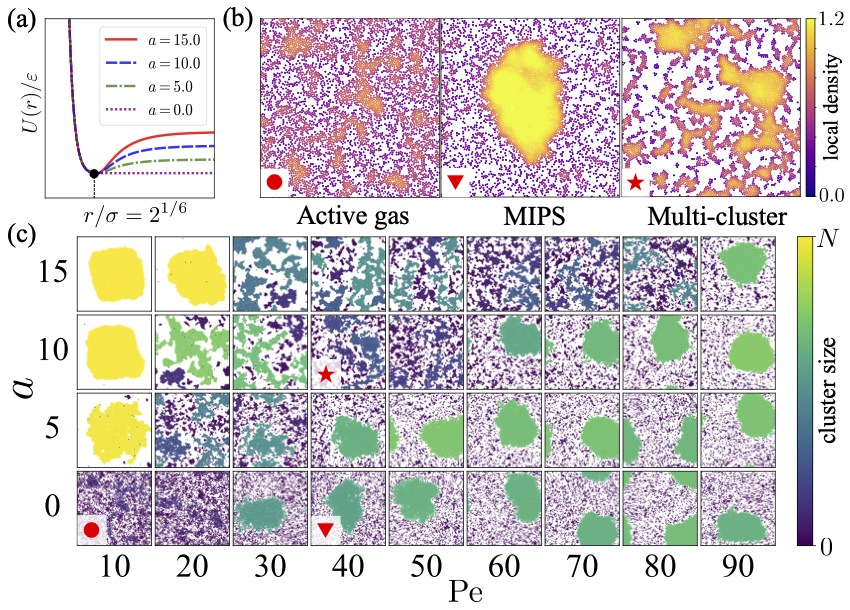}
    \caption{Active Brownian particle (ABP) model with attractive force. 
    (a)~Potential form of $U(r)$ at several values of $a$.
    (b)~Snapshots from the simulation at the parameter sets indicated by the corresponding symbols in (c). Color indicates local density.
    (c)~Phase diagram of the model, illustrated with snapshots from the simulations. Colors represent cluster size, where clusters are identified based on a distance threshold of less than $1.13\sigma$.}
    \label{fig1}
  \end{figure}

$\bm F_i = -\sum_{j\neq i} {\bm \nabla}_i U(r_{ij})$ in Eq.~\eqref{eq:EqR} represents the force acting between the particles, where $U(r)$ denotes the potential function and $r_{ij} \equiv |{\bm r}_{i}-{\bm r}_{j}|$ is the distance between the $i$-th and $j$-th particles. 
Since we are interested in attractive interactions, we introduce the modified Lennard--Jones (LJ) potential as follows:
\begin{align*}
    U(r)=
    \begin{dcases}
        \ 4E\left[ \left(\frac{\sigma}{r}\right)^{12}\!-\!\left(\frac{\sigma}{r}\right)^{6}  \right] &(r\leq2^{1/6}\sigma)\\
    4a E\left[ \left(\frac{\sigma}{r}\right)^{12}\!-\!\left(\frac{\sigma}{r}\right)^{6} \right]+E(a-1) & (r>2^{1/6}\sigma).
    \end{dcases}
\end{align*}
Here, $E$ is a parameter that defines the energy scale, similar to that of the standard LJ potential. $U(r)$ includes an additional dimensionless parameter, $a$, which controls the attraction strength while maintaining the core repulsion part of $r \leq 2^{1/6}\sigma$ unchanged, differently from previous studies~\cite{Sarkar2021}. 
Figure~\ref{fig1}(a) shows the functional form of $U(r)$.
The potential is purely repulsive at $a=0$, and becomes attractive as $a$ increases. 
The particle diameter $2^{1/6}\sigma$ remains constant and $U(r)$ is a smooth and continuous function for all values of $a$.

We conducted numerical simulations in an $L \times L$ square domain under periodic boundary conditions.
The total number of particles was fixed at $N=5,000$ and the system size at $L=100$ unless otherwise mentioned. 
Parameters $m=1.0$, $\gamma = 10.0$, $\tau_\theta=1000/3\simeq333.3$, $k_{\rm B}T = 10^{-4}$, $E= 1/24$, and $\sigma = 1.0$ were also fixed. Thus, the packing area fraction $\bar{\rho} = \pi (2^{-5/6}\sigma)^2 N/L^2\simeq0.49$ was also fixed in this study. 
The system was in an overdumped regime because $m/\gamma \ll\tau_\theta, \tau_R$.
At the initial time $t=0$, all particles were positioned within a single cluster located at the center of the domain, and the self-propelled directions $\theta_i$ were selected from a uniform distribution.
We investigated the particle dynamics by changing the motility $\rm{Pe}$ and adhesion strength $a$. 
The measurement was conducted by averaging $60$ snapshots from the simulation data, taken at every $500$ time interval after $t = 10{,}000$, unless otherwise mentioned.
See the Supplemental Material (SM) S1 for details of the implementation.

\textit{Results:}
Figure~\ref{fig1}(b) shows representative snapshots from the numerical simulations, colored according to the local density, for the three parameter sets indicated by symbols in the phase diagram shown in Fig.~\ref{fig1}(c) (see SM S3 for the determination of local density).
Without attraction ($a=0$), the ABPs transition from an active gas state (circle; Movie 1) to an MIPS state (triangle; Movie 2) as motility increased. 
The introduction of attraction from the MIPS state leads to cluster breakup, contrary to the intuitive expectation that attractive forces would enhance cluster formation. 
This results in a multi-cluster state, where clusters of various sizes, with local densities reaching up to $\rho \simeq 0.907$ and comparable to hexagonal packing by adhesion, dynamically move and exchange particles over time (star; Movie 3).
Figure~\ref{fig1}(c) shows the phase diagram illustrated by the simulation snapshots across a wide parameter range of ${\rm Pe}$ and $a$, colored according to the size of each cluster (see SM S3 for the determination of the cluster size).
With low motility (${\rm Pe}$) and a high attractive force ($a$), the particles aggregate into a cohesive cluster owing to adhesion, which we term the aggregation state (see Movie 4).
As motility increases, this cluster breaks up, leading to the emergence of a multi-cluster state. 
The phase diagram shows that the multi-cluster state arises in the parameter region between the aggregation and MIPS states through phase transition, emerging from the competition between the attractive and self-propulsion forces.

\begin{figure}[hbpt]
   \includegraphics[width=\columnwidth,bb=0 0 850 340]{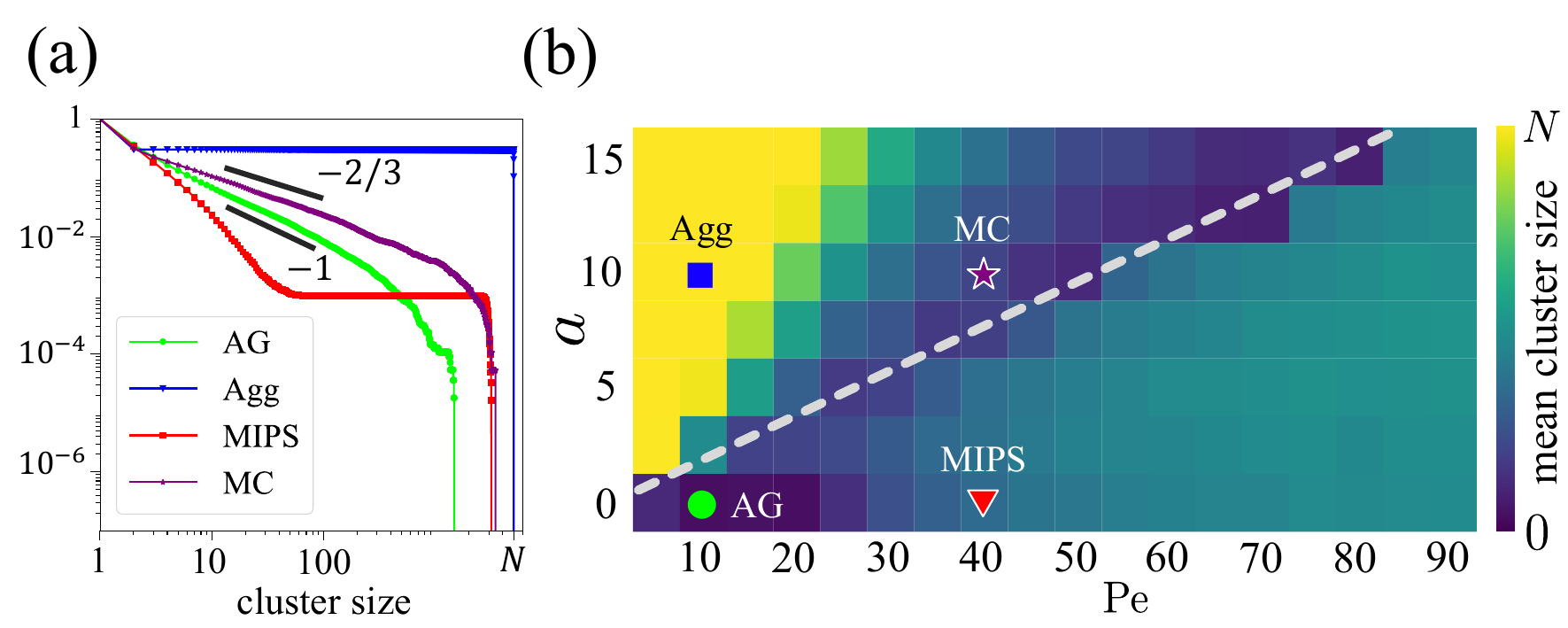}
    \caption{
    (a)~Cumulative cluster size distribution $1-P_s$ for the active gas (AG), aggregation (Agg), motility-induced phase separation (MIPS), and multi-cluster (MC) states, with each representative parameter set $(\mathrm{Pe}, a) = (10, 0), (10, 10), (40, 0), (40, 10)$.
    (b)~Mean cluster size $\overline{s} = \sum_{s=1} s^2c_s/N$. 
    Four markers represent the parameter sets used in (a). The broken gray line indicates the relation $0.19\mathrm{Pe}< a$, the condition for persistent paring of two particles (see the text).
    }
    \label{fig2}
  \end{figure}

The multi-cluster state exhibits complex dynamics involving large fluctuations and clusters of various sizes (see SM S3 for characterization based on local particle density).
To analyze this state, we measured the number of clusters of size $s$, denoted by $c_s$, and calculated the normalized frequency $p_s = c_s/M$, where $M = \sum_{s=1}^{N} c_s$ represents the total number of clusters.
Figure~\ref{fig2}(a) shows the cumulative cluster size distribution $1-P_s=1-\sum_{s'=1}^{s}p_{s'}$ for representative parameter sets.
Whereas the MIPS and aggregation states show steep drops at large $s$ reflecting the presence of a large single cluster, the active gas and multi-cluster states exhibit a power-law distribution, followed by an exponential decay.
The power-law exponent was approximately $-1$ for the active gas state. As attraction and motility increase, the exponent shifts to $-2/3$ in the multi-cluster state (see SM S3 and Fig.~S4 (a) in detail), where clusters of various sizes coexist, with the largest being comparable to a single cluster in the MIPS state.
In the multi-cluster state, the tail of the cluster size distribution evolves with the system size (see SM S3 and Fig.~S4 (b)), implying that the exponential cutoff arises from finite-size effects and that the multi-cluster state is reminiscent of critical phenomena. 
Note that the exponent $-2/3$ is consistent with the prediction of the aggregation-fragmentation process discussed in~\cite{BenNaim2008}, which states that the balance between the rate of aggregation and fragmentation leads to critical behavior.
Figure~\ref{fig2}(b) shows the mean cluster sizes $\overline{s}$ for ${\rm Pe}$ and $a$.
Here, $\overline{s}$ is given by $\overline{s} = \sum_{s=1}^{N} s^2c_s/N$ because $sc_s$ represents the total number of particles contained in clusters of size $s$ (note the identity $\sum_{s=1}^{N} sc_s = N$). 
Along with the local density fluctuations (in SM S3 and Fig.~S1), the phase diagram determines the boundaries between states. Attractive ABPs show phase transitions from the MIPS and aggregation to the multi-cluster state, and our analysis indicates that the active gas and multi-cluster states are explained by crossover, as explained below.

Why is MIPS disrupted by attraction? 
To address this, we consider a pair of ABPs in contact.
The condition for the $i$-th and $j$-th particles to remain attached is given by (see SM S2 for details):
\begin{align}
    \label{eq:StableCondition}
    f\left|\sin\left(\frac{\theta_i-\theta_j}{2}\right)\right| < \frac{504 aE}{169\sigma}\left(\frac{7}{26}\right)^{1/6} \simeq 2.4aE/\sigma~,
\end{align}
where we neglected the small noise.
Because $\theta_i$ and $\theta_j$ are uncorrelated in ABPs, the expected value of $\left|\sin(\theta_i-\theta_j)/2)\right|$ is %calculated as 
$2/\pi$.
Substituting this expectation along with other fixed parameter values yields the relationship $0.19\mathrm{Pe} < a$, which is indicated by the broken gray line in Fig.~\ref{fig2}(b). 
This line asymptotically coincides with the boundary between the multi-cluster and MIPS regions as ${\rm Pe}$ increases, suggesting the following scenario: above the gray line, particles are no longer free but move in adhesion with others, leading to a reduction in effective motility as the motility force is averaged.
This reduction in motility prevents the system from sustaining a single large cluster in MIPS. This scenario is also supported by the measurement of velocity distribution (see SM S3 and Fig.~S5).
Furthermore, for a small ${\rm Pe}$, this line is presumed to provide a crossover boundary that distinguishes the active gas from the multi-cluster state. 

To characterize the dynamics of clusters, particularly in multi-cluster states, we measured the velocity of clusters of size $s$, defined as ${\bm V}_s = (1/s)\sum_{i\in S} {\bm v}_i$, with ${\bm v}_i \equiv d{\bm r}_i/dt$, 
where $S$ denotes the set of particle indices that constitute the cluster of interest.
Because the system is nearly overdamped, the cluster velocity can be approximated as ${\bm V}_s \simeq  (f/\gamma s) \sum_{i\in S} {\bm n}_i$, where a small noise term is neglected.
The squared velocity $V_s^2$ is approximated as
\begin{align}
    \label{eq:ClusterVelocity}
    V_s^2  \simeq \frac{f^2}{\gamma^2s^2}\left[ s+\sum_{\substack{i\neq j,~i,j\in S}}\cos{(\theta_i-\theta_j)} \right]~.
\end{align}
Because the directions of particles, $\theta_i$, are mutually independent among the ABPs, one might expect that the mean value of $V_s^2$ is simply given by $\langle V_s^2 \rangle_0 = f^2/\gamma^2s$, where $\langle~\!\!\cdot~\!\!\rangle_0$ represents the average under this assumption. 

\begin{figure}[tbhp]
\includegraphics[width=\columnwidth,bb=0 0 850 907]{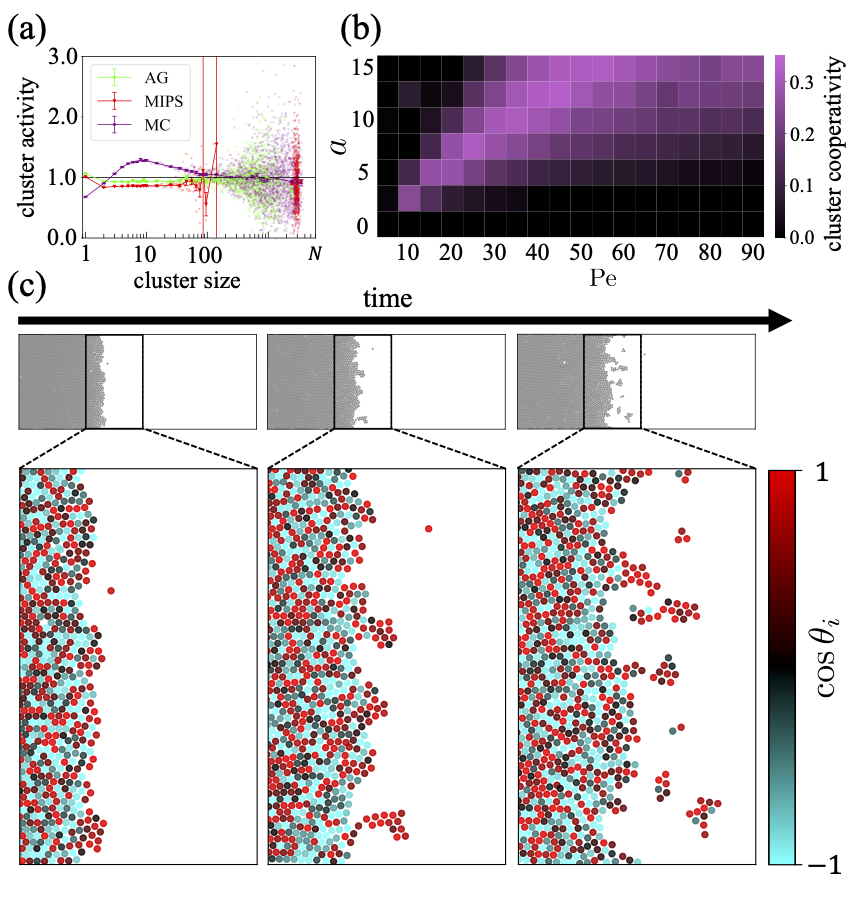} 
    \caption{(a)~Plot of cluster activity $V_s/\sqrt{\langle V_s^2\rangle_0}$, averaged over each cluster size. The corresponding parameter sets $({\rm Pe}, a)$ are: AG $(10, 0)$; MIPS $(40, 0)$; MC $(40, 10)$.
  To capture rare large clusters, we increased sampling: values were calculated from a total of 1440 snapshots, extracted every 500 time units from $t = 10{,}000$ to $t = 100{,}000$ across 8 independent simulations. 
For $s \geq 11$, the connected points with error bars represent smoothed values in binned ranges $[11,20], [21,30], \ldots, [91,100], [101,200], \ldots, [4001,5000]$. 
    (b)~Cooperativity $\Gamma_s$, averaged over all observed clusters for each parameter set with $10\leq \mathrm{Pe}\leq 90$ and $0\leq a \leq 15$.
    (c)~Illustration of cluster fragmentation in a rectangular domain in the multi-cluster state at $(\mathrm{Pe}, a) = (40, 10)$. We set $L_X=250$, $L_Y=50$, and $N=6,250$ so that the packing area fraction remains $\bar{\rho}\simeq0.49$. The particles in enlarged views are colored by $\cos \theta_i$. Red (blue) indicates particles exerting their self-propelled force toward the right (left).
    }
    \label{fig3}
  \end{figure}

However, contrary to this prediction, we found that $V_s^2$ systematically deviates from $\langle V_s^2 \rangle_0$ in the multi-cluster state---a key finding of this study.
Figure~\ref{fig3}(a) shows plots of cluster activity, defined as $V_s/\sqrt{\langle V_s^2\rangle_0}$, against cluster size $s$.
The mean values and their standard errors are shown as points and error bars, respectively. 
In the multi-cluster state, small clusters consisting of several to fewer than a hundred particles move at significantly higher velocities than anticipated, whereas in the other states, the observed velocity remains close to the expected value for randomly oriented particles. 

As expressed in Eq.~\eqref{eq:ClusterVelocity}, deviation from the expected value of unity arises from the term $\sum_{i\neq j}\cos{(\theta_i-\theta_j)}$. In other words, the deviation indicates that the particles in a cluster are correlated in their directions $\theta_i$, even though each $\theta_i$ is driven by an independent noise.
We define cooperativity of a cluster as $\Gamma_s=\sum_{i\neq j}\cos{(\theta_i-\theta_j)}/(s^2-s)$, which attains a maximum value of 1 when all particles are perfectly aligned and becomes zero when the directions are randomly distributed. Figure~\ref{fig3}(b) shows the cluster cooperativity $\Gamma_s$ against the self-propulsion strength $\mathrm{Pe}$ and attraction strength $a$.
The region with high cooperativity coincides with the multi-cluster regime, bounded by the crossover line derived above (broken gray line in Fig.~\ref{fig2}(b)), indicating that the clusters in the state move relatively fast by aligning the self-propulsion directions.

What is the underlying mechanism behind the alignment within each cluster in a multi-cluster state?
To investigate this, we performed numerical simulations of ABPs in a rectangular domain with periodic boundary conditions, where the domain was shorter along the vertical direction. 
The simulations were initialized with particles arranged in a single vertical cluster band along the center of the domain (Fig.~\ref{fig3}(c) shows the right half of the system; see Movie 5).
This setup has been used to study the interface behavior while minimizing the influence of the interface curvature~\cite{Bialk2015}. The magnified views of the interface in Fig.~\ref{fig3}(c) show the particles colored according to their directional cosine, $\cos \theta_i$, with red (cyan) indicating particles oriented to the right (left). 
We observed that the particles directed to the right near the interface aggregated and were collectively detached from the central large cluster.
This suggests that when a subset of particles near the surface of a cluster aligns incidentally, they can collectively break away and escape from the cluster.
This behavior contrasts with that observed in the MIPS state, where the attractive force is weak and individual particles escape independently without requiring collective motion (see Movie 6).

%%%%%%%%%%%%%%%%%%%%%%%%%%%%%%%%
\textit{Discussion:}
We demonstrated that ABPs with attractive interactions suppress MIPS and induce a transition to a multi-cluster state characterized by a power-law distribution of cluster sizes.
Additionally, we found that small clusters in this state exhibit enhanced mobility because of the emergent alignment of self-propulsion directions, which facilitates their escape from larger clusters.

The power-law distribution of the cluster size observed in the multi-cluster state can be understood using the framework of aggregation-fragmentation dynamics~\cite{Ginot2018}. 
Existing theory predicts that the cluster size distribution follows a power law with an exponent of $-5/3$ (or $-2/3$ for the cumulative distribution) when the rates of fragmentation and aggregation are balanced~\cite{BenNaim2008}, which is consistent with our results.
This agreement is plausible given that increased active forces promote cluster fragmentation~\cite{Palmiero2023}, whereas attractive interactions favor cluster aggregation.
Our simulation results suggest that this competition between motility and attraction is realized between the aggregation and MIPS states, giving rise to a multi-cluster state with various cluster sizes.
However, it should be noted that our simple model is not sufficient to fully account for experimental observations.
Several experimental studies on cell populations and bacterial systems have reported power-law distributions of cluster sizes with
different exponents~\cite{Mendes2001, Zhang2010,Peruani2012,Chen2012}, indicating that the exponent may depend on additional factors such as particle density and alignment interactions.

Crucially, we found a spontaneous alignment of the active forces within each cluster in the multi-cluster state, even in the absence of direct alignment interactions. 
In ABP models with purely repulsive interactions, the emergence of local alignment order within the single cluster of the MIPS state has been reported, despite the absence of explicit alignment mechanisms~\cite{Caprini2020, Paul2024}.
Although global alignment in the self-propulsion directions $\theta_i$ is unattainable without the alignment interaction, other studies have shown that ABPs with attractive interactions can exhibit global alignment of particle velocities~\cite{Caprini2023, Chen2025}.
The alignment observed in our study occurs locally within each cluster of the multi-cluster state and is associated with the directions of the self-propelled forces $\theta_i$, although those of particle velocities ${\bm v_i}$ also aligns accordingly. 

The local alignment is driven by the cooperative escape of cohesive particles that simultaneously leave a larger cluster.
Similar collective detachment events have been reported in epithelial~\cite{Campbell2015, Chepizhko2016, Fu2024} and bacterial~\cite{Zdimal2025} systems, where small clusters of cells separate from a larger group and migrate coherently. 
While these phenomena have been reproduced in using particle model in one dimensional ~\cite{George2017} and cellular Potts model~\cite{Mukherjee2021}, our simple model provides a minimal mechanism for the collective and rapid spread of detached particles, even in the absence of explicit alignment rules.

In conclusion, attractive interactions can play intricate and subtle roles in motile systems, not only in suppressing aggregation but also in facilitating alignment. 
The present model, along with its possible extensions, may help to elucidate how attractive forces are harnessed in biological and other active matter systems.

\vspace{1em}

This study was supported by JST SPRING JPMJSP2108 (to S.S.), JSPS KAKENHI Grant Number 24H01931 (to S.I.),  and JST CREST JPMJCR1923 (to S.I.).
We thank K. Mitsumoto and T. Namba for their fruitful discussions and kind support.

\vspace{1em}

S.S., N.S., and S.I. proposed the research direction, contributed to the theoretical analysis, and wrote the manuscript.
S.S. performed all the numerical simulations.

\bibliographystyle{junsrt}
\bibliography{citation}

\begin{thebibliography}{10}

\bibitem{Friedl2009}
Peter Friedl and Darren Gilmour.
\newblock Collective cell migration in morphogenesis, regeneration and cancer.
\newblock {\em Nature Reviews Molecular Cell Biology}, Vol.~10, No.~7, p.
  445^^e2^^80^^93457, July 2009.

\bibitem{Vedula2013}
Sri Ram~Krishna Vedula, Andrea Ravasio, Chwee~Teck Lim, and Benoit Ladoux.
\newblock Collective cell migration: A mechanistic perspective.
\newblock {\em Physiology}, Vol.~28, No.~6, p. 370^^e2^^80^^93379, November
  2013.

\bibitem{Lu2021}
Pengfei Lu and Yunzhe Lu.
\newblock Born to run? diverse modes of epithelial migration.
\newblock {\em Frontiers in Cell and Developmental Biology}, Vol.~9, , August
  2021.

\bibitem{Copeland2009}
Matthew~F. Copeland and Douglas~B. Weibel.
\newblock Bacterial swarming: a model system for studying dynamic
  self-assembly.
\newblock {\em Soft Matter}, Vol.~5, No.~6, p. 1174, 2009.

\bibitem{Zhang2010}
H.~P. Zhang, Avraham {Be'er}, E.-L. Florin, and Harry~L. Swinney.
\newblock Collective motion and density fluctuations in bacterial colonies.
\newblock {\em Proceedings of the National Academy of Sciences}, Vol. 107,
  No.~31, p. 13626^^e2^^80^^9313630, July 2010.

\bibitem{Peruani2012}
Fernando Peruani, J{\"o}rn Starru^^c3^^9f, Vladimir Jakovljevic, Lotte
  S{\o}gaard-Andersen, Andreas Deutsch, and Markus B{\"a}r.
\newblock Collective motion and nonequilibrium cluster formation in colonies of
  gliding bacteria.
\newblock {\em Physical Review Letters}, Vol. 108, No.~9, February 2012.

\bibitem{Fodor2018}
{\'E}tienne Fodor and M.~Cristina Marchetti.
\newblock The statistical physics of active matter: From self-catalytic
  colloids to living cells.
\newblock {\em Physica A: Statistical Mechanics and its Applications}, Vol.
  504, p. 106^^e2^^80^^93120, August 2018.

\bibitem{Shaebani2020}
M.~Reza Shaebani, Adam Wysocki, Roland~G. Winkler, Gerhard Gompper, and Heiko
  Rieger.
\newblock Computational models for active matter.
\newblock {\em Nature Reviews Physics}, Vol.~2, No.~4, p. 181^^e2^^80^^93199,
  March 2020.

\bibitem{Fily2012}
Yaouen Fily and M.~Cristina Marchetti.
\newblock Athermal phase separation of self-propelled particles with no
  alignment.
\newblock {\em Physical Review Letters}, Vol. 108, No.~23, June 2012.

\bibitem{Cates2015}
Michael~E. Cates and Julien Tailleur.
\newblock Motility-induced phase separation.
\newblock {\em Annual Review of Condensed Matter Physics}, Vol.~6, No.~1, p.
  219^^e2^^80^^93244, March 2015.

\bibitem{Sarkar2021}
Debarati Sarkar, Gerhard Gompper, and Jens Elgeti.
\newblock A minimal model for structure, dynamics, and tension of monolayered
  cell colonies.
\newblock {\em Communications Physics}, Vol.~4, No.~1, February 2021.

\bibitem{Caprini2023}
L.~Caprini and H.~L\"{o}wen.
\newblock Flocking without alignment interactions in attractive active brownian
  particles.
\newblock {\em Physical Review Letters}, Vol. 130, No.~14, April 2023.

\bibitem{Theurkauff2012}
I.~Theurkauff, C.~Cottin-Bizonne, J.~Palacci, C.~Ybert, and L.~Bocquet.
\newblock Dynamic clustering in active colloidal suspensions with chemical
  signaling.
\newblock {\em Physical Review Letters}, Vol. 108, No.~26, June 2012.

\bibitem{SchwarzLinek2012}
J.~Schwarz-Linek, C.~Valeriani, A.~Cacciuto, M.~E. Cates, D.~Marenduzzo, A.~N.
  Morozov, and W.~C.~K. Poon.
\newblock Phase separation and rotor self-assembly in active particle
  suspensions.
\newblock {\em Proceedings of the National Academy of Sciences}, Vol. 109,
  No.~11, p. 4052^^e2^^80^^934057, March 2012.

\bibitem{Mognetti2013}
B.~M. Mognetti, A.~{\v{S}}ari{\'c}, S.~Angioletti-Uberti, A.~Cacciuto,
  C.~Valeriani, and D.~Frenkel.
\newblock Living clusters and crystals from low-density suspensions of active
  colloids.
\newblock {\em Physical Review Letters}, Vol. 111, No.~24, December 2013.

\bibitem{Mendes2001}
Rosemairy~L. Mendes, An{\'e}sia~A. Santos, M.L. Martins, and M.J. Vilela.
\newblock Cluster size distribution of cell aggregates in culture.
\newblock {\em Physica A: Statistical Mechanics and its Applications}, Vol.
  298, No. 3^^e2^^80^^934, p. 471^^e2^^80^^93487, September 2001.

\bibitem{Chen2012}
Xiao Chen, Xu~Dong, Avraham {Be'er}, Harry~L. Swinney, and H.~P. Zhang.
\newblock Scale-invariant correlations in dynamic bacterial clusters.
\newblock {\em Physical Review Letters}, Vol. 108, No.~14, April 2012.

\bibitem{BenNaim2008}
E.~Ben-Naim and P.~L. Krapivsky.
\newblock Phase transition with nonthermodynamic states in reversible
  polymerization.
\newblock {\em Physical Review E}, Vol.~77, No.~6, June 2008.

\bibitem{Bialk2015}
Julian Bialk{\'e}, Jonathan~T. Siebert, Hartmut L{\"o}wen, and Thomas Speck.
\newblock Negative interfacial tension in phase-separated active brownian
  particles.
\newblock {\em Physical Review Letters}, Vol. 115, No.~9, August 2015.

\bibitem{Ginot2018}
F.~Ginot, I.~Theurkauff, F.~Detcheverry, C.~Ybert, and C.~Cottin-Bizonne.
\newblock Aggregation-fragmentation and individual dynamics of active clusters.
\newblock {\em Nature Communications}, Vol.~9, No.~1, February 2018.

\bibitem{Palmiero2023}
Miriam Palmiero, Isabel Cantarosso, Laura di~Blasio, Valentina Monica, Barbara
  Peracino, Luca Primo, and Alberto Puliafito.
\newblock Collective directional migration drives the formation of heteroclonal
  cancer cell clusters.
\newblock {\em Molecular Oncology}, Vol.~17, No.~9, p. 1699^^e2^^80^^931725,
  January 2023.

\bibitem{Caprini2020}
L.~Caprini, U.~Marini Bettolo~Marconi, and A.~Puglisi.
\newblock Spontaneous velocity alignment in motility-induced phase separation.
\newblock {\em Physical Review Letters}, Vol. 124, No.~7, February 2020.

\bibitem{Paul2024}
Subhajit Paul, Suman Majumder, and Wolfhard Janke.
\newblock Spontaneous micro flocking of active inertial particles without
  alignment interaction.
\newblock {\em arXiv preprint arXiv:2402.04397}, 2024.

\bibitem{Chen2025}
Jian-li Chen, Jia-jian Li, and Bao-quan Ai.
\newblock Spontaneous velocity alignment of active particles with rotational
  inertia.
\newblock {\em Physica A: Statistical Mechanics and its Applications}, Vol.
  658, p. 130279, January 2025.

\bibitem{Campbell2015}
Kyra Campbell and Jordi Casanova.
\newblock A role for e-cadherin in ensuring cohesive migration of a
  heterogeneous population of non-epithelial cells.
\newblock {\em Nature Communications}, Vol.~6, No.~1, August 2015.

\bibitem{Chepizhko2016}
Oleksandr Chepizhko, Costanza Giampietro, Eleonora Mastrapasqua, Mehdi
  Nourazar, Miriam Ascagni, Michela Sugni, Umberto Fascio, Livio Leggio, Chiara
  Malinverno, Giorgio Scita, St^^c3^^a9phane Santucci, Mikko~J. Alava, Stefano
  Zapperi, and Caterina A.~M. La~Porta.
\newblock Bursts of activity in collective cell migration.
\newblock {\em Proceedings of the National Academy of Sciences}, Vol. 113,
  No.~41, p. 11408^^e2^^80^^9311413, September 2016.

\bibitem{Fu2024}
Chaoyu Fu, Florian Dilasser, Shao-Zhen Lin, Marc Karnat, Aditya Arora, Harini
  Rajendiran, Hui~Ting Ong, Nai Mui Hoon~Brenda, Sound~Wai Phow, Tsuyoshi
  Hirashima, Michael Sheetz, Jean-Fran\c{c}ois Rupprecht, Sham Tlili, and
  Virgile Viasnoff.
\newblock Regulation of intercellular viscosity by e-cadherin-dependent
  phosphorylation of egfr in collective cell migration.
\newblock {\em Proceedings of the National Academy of Sciences}, Vol. 121,
  No.~37, September 2024.

\bibitem{Zdimal2025}
Amanda~M Zdimal, Giacomo Di~Dio, Wanxiang Liu, Tanya Aftab, Taryn Collins, Remy
  Colin, and Abhishek Shrivastava.
\newblock Swarming bacteria exhibit developmental phase transitions to
  establish scattered colonies in new regions.
\newblock {\em The ISME Journal}, Vol.~19, No.~1, January 2025.

\bibitem{George2017}
Mishel George, Francesco Bullo, and Otger Camp{\`a}s.
\newblock Connecting individual to collective cell migration.
\newblock {\em Scientific Reports}, Vol.~7, No.~1, August 2017.

\bibitem{Mukherjee2021}
Mrinmoy Mukherjee and Herbert Levine.
\newblock Cluster size distribution of cells disseminating from a primary
  tumor.
\newblock {\em PLOS Computational Biology}, Vol.~17, No.~11, p. e1009011,
  November 2021.

\end{thebibliography}

\end{document}